\documentclass[11pt,twoside]{article}

\usepackage{asp2006}
\usepackage{graphicx}

\begin{document}

\title{Probing the History of Cluster Assembly by Intracluster Planetary Nebulae}
\author{Sadanori Okamura}   
\affil{Department of Astronomy and Research Center for the Early Universe,
School of Science, the University of Tokyo, Tokyo, 113-0033 Japan}

\begin{abstract} 
We describe the results and the current status of our program to search
for intracluster planetary nebulae (ICPNe) in the Virgo and the Coma
clusters of galaxies. For the Virgo cluster, ICPNe candidates were
detected with narrow band imaging combined with
broad band imaging. The use of two narrow band
filters for H$\alpha$ and [O III] lines tuned for the cluster redshift
enabled us to detect secure ICPNe candidates. Kinematics of the detected
ICPNe confirm that the Virgo cluster is in a highly unrelaxed state.
In order to detect ICPNe in the Coma cluster, which is more than five times
more distant than the Virgo cluster, we devised a multi-slit imaging
spectroscopy technique and successfully detected 35 secure ICPNe
candidates for the first time in this cluster.
We have found a hint that the Coma cluster is currently in the
midst of a subcluster merger.
\end{abstract}


\section{Introduction}

Cosmological simulations predict that a substantial fraction of stars
in galaxies were stripped off during the assembly of a cluster of
galaxies. These
stars were dispersed into intracluster space and now recognized as
the intracluster stellar population. This population is a sensitive
probe of cluster assembly, and more generally, structure formation
of the universe. The total amount, spatial distribution (morphology),
kinematics, and metallicty are among the key parameters which
characterize the population.

The intracluster stellar population has been observed as diffuse
intracluster light \citep{Mihos05,Krick06},
or as individual stars, i.e., planetary nebulae \citep{Arnaboldi96,Okamura02},
or as red giant stars \citep{Ferguson98}.
The intracluster planetary nebulae (ICPNe) are
the only probe that enables us to investigate the kinematics
of the intracluster stellar population.
We launched a program to search for ICPNe in clusters of
galaxies with a range of properties on the advent of Subaru
telescope and its wide-field Camera, Suprime-Cam.

\section{Observing Strategy}

Initial targets of our program are the Virgo cluster at the
distance of $(m-M)_0\sim31$ and the Coma cluster at $(m-M)_0\sim35$.
The Virgo cluster is irregular and loose in shape, and dynamically
unrelaxed containing a considerable fraction of spiral galaxies, while
the Coma cluster is regular and rich in shape, and
dynamically more relaxed than the Virgo cluster.

We applied a conventional narrow band plus broad band imaging
technique to the nearby Virgo cluster.
Unique feature of our search is the use of two narrow band
filters for H$\alpha$ and [O III] lines tuned for the
cluster redshift. Previous planetary nebula searches over the
image of Virgo ellipticals \citep[e.g,][hereafter JCF]{Jacoby90}
used the strongest [O III] line only. The PNe candidates
selected by the [O III] single line technique is contaminated
by foreground and background emission line objects.
Especially, when the single line technique is used to search
for ICPNe in the 'blank region' free from Virgo member
galaxies, the contamination could dominate the candidates.
Use of the two lines and broad band images enabled us
to identify secure ICPNe candidates. However, the follow-up
spectroscopy is necessary to the final confirmation
and kinematic studies. We used Telescopio Nazionale Galileo (TNG)
4 m telescope and Very Large Telescope (VLT) for
the follow-up spectroscopy.

No imaging technique can be successfully applied to the
detection of ICPNe in the distant Coma cluster.
At the Coma distance, even the
brightest PN is buried in the sky noise and undetectable
in the narrow band image.
Increase of the signal-to-noise
ratio of the emission lines by reducing the sky noise
is critically important. To achieve this, we devised a
multi-slit imaging spectroscopy (MSIS) technique, which
was sucessfully applied to the Coma cluster.

\section{Virgo Cluster}

\subsection{Imaging Observation and Selection Criteria}

Observations, selection criteria for secure PNe candidates,
and initial results were described in \citet{Okamura02}
and \citet{Arnaboldi03}.
Only the outline is repeated here.

\begin{figure}[h!]
\begin{center}
\includegraphics[width=13cm]{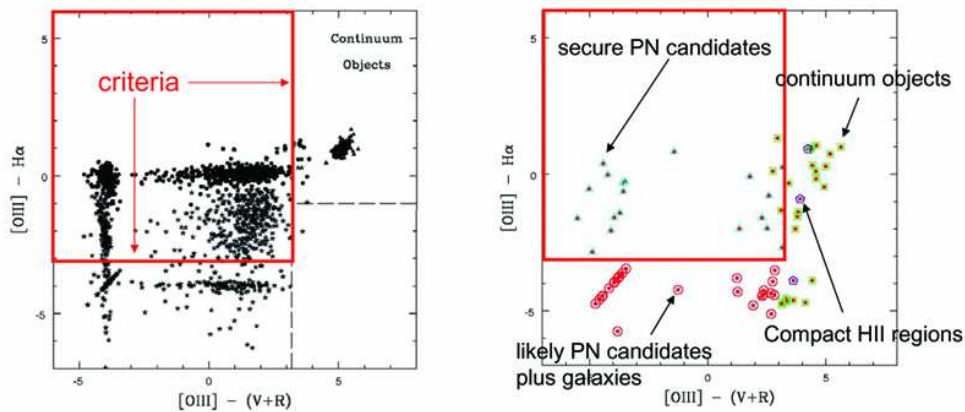}
\caption{\small{(left) Two-color diagram of the simulated population
of point sources. Dots are PNe, stars are single emission-line objects,
and triangles are continuum objects. (right) Two-color diagram for
the M84 matched sample. Dots are [O III] sources matched to the JCF
sample. Triangles (19 objects) indicate the best PN candidates, and
circles (26 objects) indicate the single-line emitters, most likely
high excitation PNe in this situation. Squares indicate continuum objects.
Three pentagons indicate objects with unreliable colors, and one of
them was found to be a compact H II region apparently floating
in the intracluster space \citep{Gerhard02}.
Our selection criteria are shown by the thick lines.
\citep{Arnaboldi03}}}
\end{center}
\label{okamura_Fig1}
\end{figure}

In 2001 March - April a field in the Virgo Cluster at
($\alpha$, $\delta$)(J2000) = ($12^h25^m47.^s0$,
+12$^\circ$43$'$58$''$), just south of
M84-M86, was observed during the commissioning of the
Suprime-Cam \citep{Miyazaki02}, at the prime focus
of the Subaru 8.2 m telescope \citep{Iye04}.
Suprime-Cam covers an area of 34$'\times27'$ of the sky,
with a resolution of 0.$^{''}$2 pixel$^{-1}$.
The field was imaged through two narrow band and two
broadband ($V$ and $R$) filters. The two narrow band filters have
($\lambda_{\rm c}$, $\Delta\lambda$)  =
(5021\AA, 74\AA) and (6607\AA, 101\AA), corresponding
to the [O III] and H$\alpha$ emissions at the redshift of
the Virgo cluster. Exposure times are 900, 720, 3600, and 8728
seconds for $V, R,$ [O III], and H$\alpha$, respectively.

We use the two-color diagram [O III] - H$\alpha$ versus
[O III] - (V+R) to select our emission-line objects.
Simulated populations of PNe, single-lined emitters, and
continuum sources were constructed and used to
outline the regions in the two-color diagram inhabited by
the different kinds of objects (Fig. 1, left panel).
On the basis of this simulation,
we set selection criteria. Two bright galaxies M86
(NGC 4406) and M84 (NGC 4374) appear in our field.
These two galaxies were surveyed for PNe by JCF,
and we can evaluate our selection criteria against the
JCF photometry of matched samples.
The right panel of Fig. 1 shows the comparison for
M84 field, where 74 objects
are matched. We found that at least 39\% contamination is
present in the JCF sample selected by [OIII] line only.

\subsection{Follow-up Spectroscopy for the Virgo Cluster}

We carried out follow-up spectroscopy of the ICPN
candidates with multifiber FLAMES spectrograph at UT2 on VLT
\citep{Arnaboldi04}. Candidates were drawn from the
three survey fields, FCJ, CORE, and SUB, as shown in
the left panel of Fig.2.
Spectra covered a wavelength range of 500\AA, centered on
4797\AA~with a resolution of $\lambda/\Delta\lambda=7500$.
This gives a velocity resolution of
40 km s$^{-1}$ and a typical velocity error of 12 km s$^{-1}$.

A large fraction of the photometric candidates
with m(5007)$<$27.2 were confirmed and $\lambda$4959
line of the [O III] doublet was detected for the first time
in about a half of confirmed ICPNe. A total of 40 ICPNe candidates
(15/12/13 for FCJ/CORE/SUB fields) were confirmed, and their
radial velocities were measured.

\begin{figure}[h]
\begin{center}
\begin{minipage}{6cm}
\includegraphics[width=6cm]{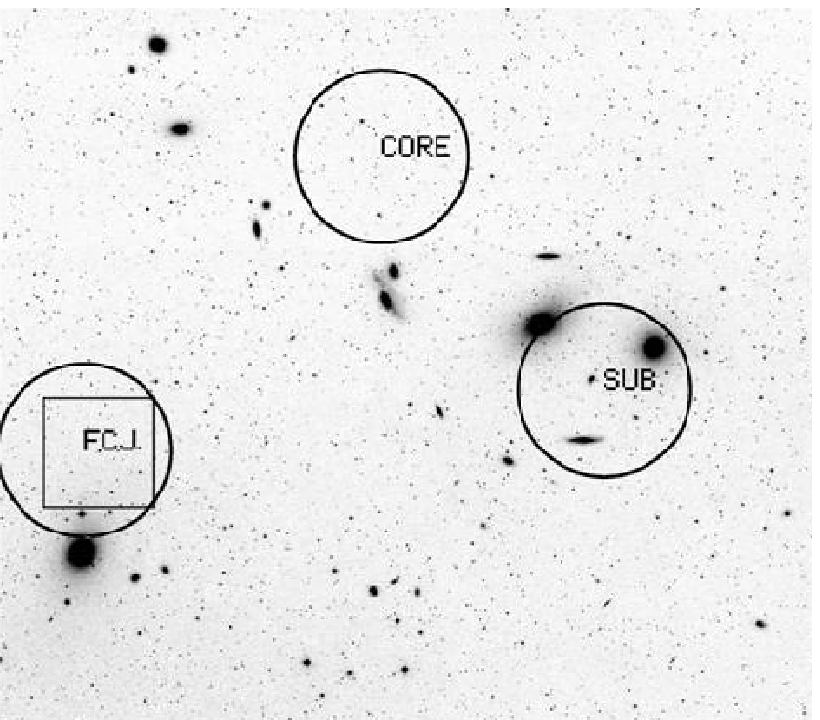}
\end{minipage}
\begin{minipage}{6cm}
\includegraphics[width=6cm]{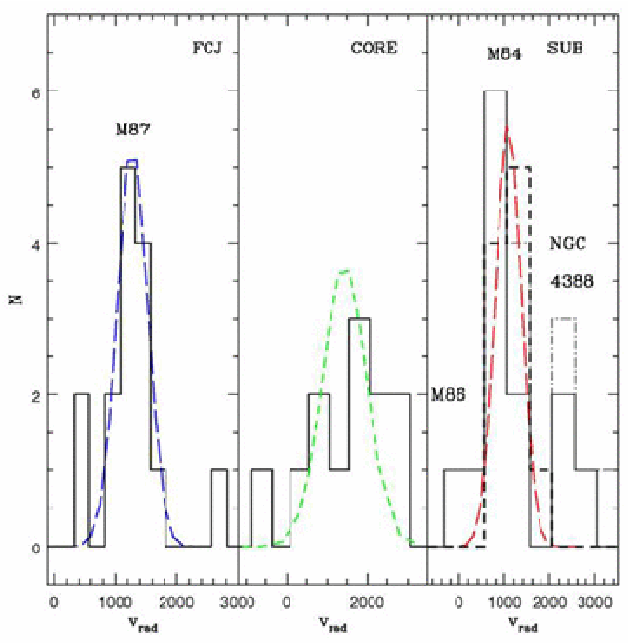}
\end{minipage}
\caption{\small{(left) Three fields in the Virgo cluster
where follow-up spectroscopy was made. Circles show FLAMES
field of view. (right) Radial velocity distributions
of ICPNe in the three fields. \citep{Arnaboldi04}}}
\end{center}
\label{okamura_Fig2}
\end{figure}

Radial velocity distributions of the ICPNe in the three fields
are shown in the right panel of Fig. 2. They are considerably
different from each other. This confirms the view that Virgo
is a highly nonuniform and unrelaxed galaxy cluster, consisting
of several subunits that have not yet had time to come to
equilibrium in a common gravitational potential.

\section{Toward the Coma Cluster}

\subsection{MSIS Technique}

As already mentioned, no imaging technique can be successfully
applied to the detection of PNe in the Coma cluster.
We devised a technique that is similar to the
approach used to search for Ly$\alpha$-emitting galaxies at very high
redshift \citep{Stern99}.
It combines multi-slit grism spectroscopy with a narrow band filter
centered around the [O III] $\lambda$5007 line at the redshift
of the Coma cluster. We use the Faint Object Camera and Spectrograph
\citep[FOCAS;][]{Kashikawa02} at the Cassegrain focus of
Subaru telescope (see \citeauthor{Gerhard05} \citeyear{Gerhard05} for the method and
initial results).

A mask of parallel multiple slits is used to obtain spectra
of all PNe that happen to lie behind the slits.
Because the [O III] emission lines from PNs are only
a few kilometers per second wide, their entire flux still falls into
a small number of pixels in the two-dimensional spectrum, determined
by the slit width and seeing. On the contrary, the sky
emission is dispersed in wavelength, allowing a large increase
in the signal-to-noise ratio (S/N). The narrow band filter limits
the length of the spectra on the CCD detector, so that many slits can be
exposed. This technique, which is a blind search, is referred to as
the multi-slit imaging spectroscopy (MSIS) technique.
No conventional imaging technique
can decrease the sky surface brightness in a similar way.

\begin{figure}[h!]
\begin{center}
\includegraphics[width=10cm]{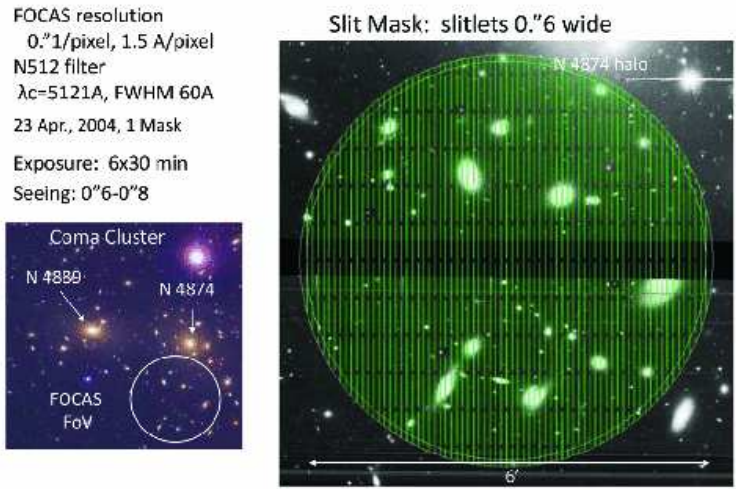}
\caption{\small{MSIS technique. (left) Location of the
MSIS field. (right) Slit mask. \citep{Gerhard05}}}
\end{center}
\label{okamura_Fig3}
\end{figure}

Figure 3 shows the MSIS field and the slit mask.
The circular FOCAS field of view
(FOV) has $6'$ in diameter (166 kpc at the assumed distance of
95 Mpc). The spatial resolution of FOCAS is $0.''1$ pixel$^{-1}$, so
the $6'$ diameter corresponds to 3600 pixels.
The light passing through the narrow band filter and a $0.''6$ wide
long slit, and dispersed by the grism, projects down to a spectrum
of about 43 pixels. A mask was therefore manufactured with uniform
long slits spaced every 50 pixels and interrupted only by
short sections to ensure mechanical stability.
We used 300B grating, which gives a measured
dispersion of 1.45\AA~pixel$^{-1}$ on the two FOCAS CCD chips.
The effective spectral resolution is 7.3\AA , or 440 km s$^{-1}$.
We also used the N512 filter with FWHM 60\AA, centered at
$\lambda_c$ = 5121\AA , the wavelength of the redshifted [O III] emission
from a PN at the mean recession velocity of the Coma Cluster
\citep[6853 km s$^{-1}$;][]{Colless96}. The 60\AA~FWHM
includes only $\pm1.6\sigma$ the velocity dispersion of galaxies.
By taking exposures at 8 positions sucessively shifted
along the dispersion direction, we can roughly cover the
whole field of view.

\subsection{Hint of On-going Subcluster Merger in the Coma Cluster}

Observations at three mask positions centered on the field at
($\alpha, \delta$) (J2000) = (12$^h$59$^m$41.$^s$8,
+27$^\circ$53$'$25.$''$4) (see Fig. 3) were
carried out during the nights of 2004 April
21-23 . We detected a total of 60 emission line objects.
Among them, 35 are ICPNe candidates with no continuum flux,
5 are PN candidates projected against the extended halos of
Coma member galaxies, and 20 are background galaxies.
Details of the data reduction and a catalog of ICPNe candidates
are given in \citet{Arnaboldi07}.

\begin{figure}[h!]
\begin{center}
\includegraphics{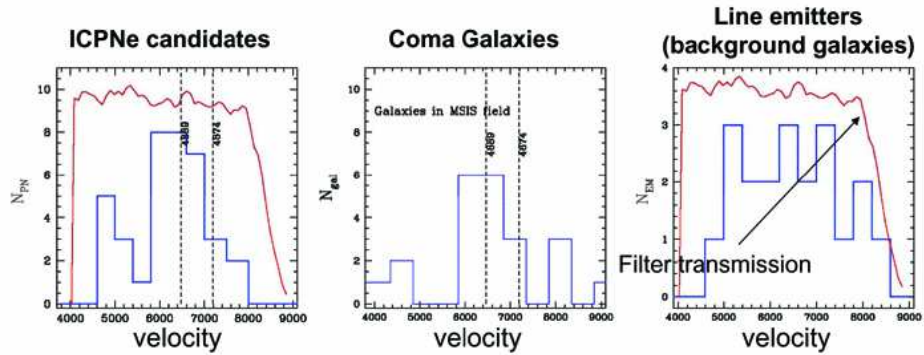}
\caption{\small{Velocity histograms of ICPNe (left),
Coma galaxies in the MSIS field (middle), and background
emission-line galaxies (right). Filter transmission curve
is shown in the left and right panels. (Gerhard et al. 2007)}}
\end{center}
\label{okamura_Fig4}
\end{figure}

Figure 4 shows the velocity histograms of ICPNe (left),
Coma galaxies in the MSIS field (middle), and background
emission-line galaxies (right).
We detect clear velocity substructures of ICPNe, i.e.,
intracluster stellar population, within a 6 arcmin
diameter MSIS field. A substructure is present at $\sim5000$ km
s$^{-1}$, which is probably from infall of a galaxy group,
while the main component is centered around $\sim6500$ km s$^{-1}$,
$\sim700$ km s$^{-1}$ offset from the nearby cD galaxy NGC 4874.
The kinematics of ICPNe found in this study and
morphology of the diffuse intracluster
light \citep{Thuan77} show that the cluster
core is in a highly dynamically evolving state. In combination with galaxy
redshift and X-ray data this argues strongly that the cluster
is currently in the midst of a subcluster merger during which the
elongated distribution of diffuse light has been created.
The two subcluster cores (NGC 4874 and NGC 4889) are presently
just after their second close passage. The likely orbits of
the subcluster cores are illustrated in Fig. 5. \citep{Gerhard07}

\begin{figure}[h!]
\begin{center}
\includegraphics[width=10cm]{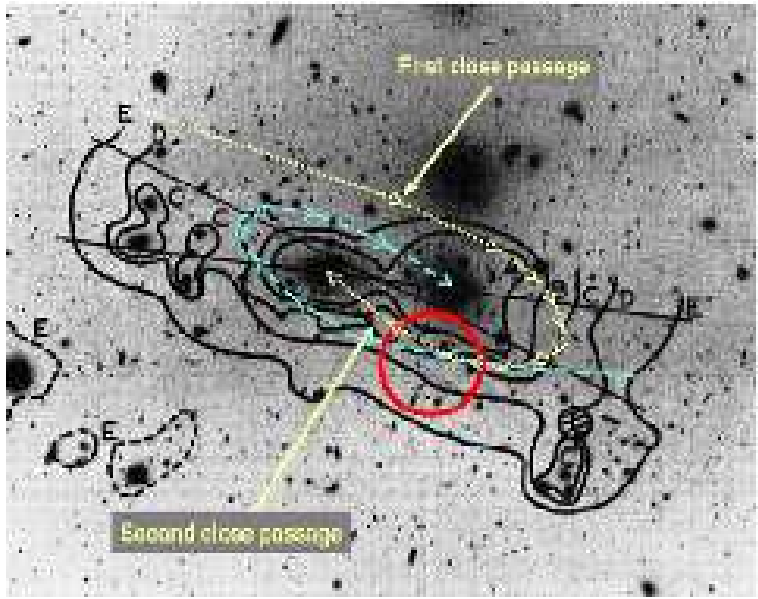}
\caption{\small{The likely orbits
of NGC 4889 and NGC 4874 up to their present positions are
sketched. Contours show the surface brightness distribution
of diffuse intracluster light taken from
\citet{Thuan77}. The circle indicates
the MSIS field. \citep{Gerhard07}}}
\end{center}
\label{okamura_Fig5}
\end{figure}

\subsection{New Deep Imaging Program for Diffuse Intracluster Light}

Inspired by the work by \citet{Mihos05} and \citet{Krick06},
we started a new very deep imaging program to detect the diffuse
intracluster light in a wide $50'\times50'$ field centered on
the Coma cluster using a CCD camera attached to the 105-cm Schmidt
telescope at Kiso observatory, Institute of Astronomy,
University of Tokyo. We hope to obtain some data in spring, 2008.

\vspace{1truecm}

\acknowledgements 

This work is based on the collaboration with Magda Arnaboldi, Ortwin Gerhard,
Ken Freeman, Naoki Yasuda, and Nobunari Kashikawa, and in part with
Suprime-Cam team.
We thank the staff of Subaru Observatory for their support.


\vspace{1truecm}
{\bf Questions and Answers}\\ \\

{\it M. Strauss}: How large would the contamination of the
Virgo ICPNe sample be by background galaxies if you didn't have the
H$\alpha$ filter?

{\it S.Okamura}: Lower limits of the fraction of contamination estimated
from the comparison with FCJ are 39\% and 56\% for M84 sample and
M86 sample, respectively. In a blank intracluster field,
the contamination could be much larger. So, the use of H$\alpha$ filter
is critically important.\\

{\it N. Kawai}: Can you tell whether these ICPNe were originally
formed in the member galaxies or formed outside?

{\it S.Okamura}: Most of them are likely to be formed in the
member galaxies and stripped off later during the cluster
assembly. It is, however, still an open question whether or not
some ICPNe were formed outside galaxies in the very early
phase of cluster formation.\\

{\it J. Krick}: G. Bernstein et al. measured the amount of diffuse
intracluster light in the Coma cluster to be $\sim50$\% of the
total cluster light. How do your values from the ICPNe compare to this?

{\it S.Okamura}: We have not yet made the estimate for the Coma
cluster. It is probably highly unreliable if we make the estimate
based on the number of ICPNe detected in the present study
over a fraction ($\sim$3/8) of a single small MSIS field. 
I hope our new deep imaging program with
Kiso Schmidt telescope would give a reliable estimate.\\

{\it N. Tamura}: Any estimation of metallicity for ICPNe?

{\it S.Okamura}: Estimation of metallicity is a very difficult job
yet to be done.\\


\begin{thebibliography}{}
\bibitem[Arnaboldi et al.(1996)]{Arnaboldi96}
Arnaboldi, M. et al. 1996, \apj, 472, 145
\bibitem[Arnaboldi et al.(2003)]{Arnaboldi03}
Arnaboldi, M. et al. 2003, \aj, 125, 514
\bibitem[Arnaboldi et al.(2004)]{Arnaboldi04}
Arnaboldi, M. et al. 2004, \apjl, 614, L33
\bibitem[Arnaboldi et al.(2007)]{Arnaboldi07}
Arnaboldi, M. et al. 2007, \pasj, 59, 419
\bibitem[Colless \& Dunn(1996)]{Colless96}
Colless, M. \& Dunn, A.W. 1996, \apj, 458, 435
\bibitem[Ferguson et al.(1998)]{Ferguson98}
Ferguson, H. et al. 1998, \nat, 391, 461
\bibitem[Gerhard et al.(2002)]{Gerhard02}
Gerhard, O. et al. 2002, \apjl, 580, L121
\bibitem[Gerhard et al.(2005)]{Gerhard05}
Gerhard, O. et al. 2005, \apjl, 621, L93
\bibitem[Gerhard et al.(2007)]{Gerhard07}
Gerhard, O. et al. 2007, \aap, 468, 815
\bibitem[Iye et al.(2004)]{Iye04}
Iye et al. 2004, \pasj, 56, 381
\bibitem[Jacoby et al.(1990)]{Jacoby90}
Jacoby, G. et al. 1990, \apj, 356, 332
\bibitem[Kashikawa et al.(2002)]{Kashikawa02}
Kashikawa et al. 2002, \pasj, 54, 819
\bibitem[Krick et al.(2006)]{Krick06}
Krick et al. 2006, \aj, 131, 168
\bibitem[Mihos et al.(2005)]{Mihos05}
Mihos et al. 2005, \apjl, 631, L41
\bibitem[Miyazaki et al.(2002)]{Miyazaki02}
Miyazaki, S. et al. 2002, \pasj, 54, 833
\bibitem[Okamura et al.(2002)]{Okamura02}
Okamura, S. et al. 2002, \pasj, 54, 883
\bibitem[Stern \& Spinrad(1999)]{Stern99}
Stern, D. \& Spinrad, H. 1999, \pasp, 111, 1475
\bibitem[Thuan \& Kormendy(1997)]{Thuan77}
Thuan, T.X. \& Kormendy, J. 1977, \pasp, 89, 466
\end{thebibliography}
\end{document}